\documentstyle[12pt]{article}\textheight 8.5 in\textwidth
6 in  \hoffset=-1.5 cm \voffset=0 cm
\def\eg{{$e.g.$\ }}
\def\ie{{$i.e.$\ }}
\def\ket#1{| #1 \rangle}
\def\bra#1{\langle #1 |}
\def\VEV#1{\langle #1\rangle}
\begin{document}
\noindent
G\"{o}teborg ITP 91-51\\
\noindent September 1991\\ \vspace*{25mm}
\begin{center}
{\large\bf {\Large\bf C}OSETS AS GAUGE SLICES IN SU(1,1) STRINGS}\\ \vspace*{30
mm} \rm STEPHEN HWANG\\
\vspace*{5 mm}
\small
\it Institute of Theoretical Physics\\
S-412 96 G\"{o}teborg, Sweden\\
E-mail: tfesh@secthf51\\
\vspace*{45 mm}
\normalsize
\rm ABSTRACT \\
\end{center}
\begin{quote}We consider a string theory based on an $SU(1,1)$
Wess-Zumino-Novikov-Witten model and an arbitrary unitary conformal
field theory. We show that the solutions of the Virasoro conditions,
in the unitarity regime of the $SU(1,1)$ theory,
are states which lie in the Euclidean coset $SU(1,1)/U(1)$. This shows the
validity, at the quantum level,  of a time-like type of gauge in these models.
\end{quote} \normalsize
%\unnumberedchapters%

\newpage \pagenumbering{arabic}\pagestyle{plain} \normalsize \rm
There has recently been a substantial interest in
Wess-Zumino-Novikov-Witten models (WZNW) based on non-compact Lie
groups, in particular, the simplest one with an $SU(1,1)$ symmetry. There
are currently two principal fields of investigations. In one of them,
the $SU(1,1)$ WZNW theory is considered as a string theory with the time
direction being one of the coordinates (the compact one) of the
$SU(1,1)$ manifold \cite{BALOG}-\cite{HHRS}. These string
theories have many interesting properties and they have been
proved to be unitary in the bosonic case \cite{HWANG} and in the $N=1$
fermionic case \cite{HH}. In addition they can be made
modular invariant \cite{HHRS,HHRS2} by introducing
new sectors of states, so called "non-Abelian" winding
sectors.

The other field of investigation has been focusing on taking cosets of
the $SU(1,1)$ WZNW
model \cite{LYKKEN}-\cite{DISTLER}. The resulting theories have attracted
particularly much attention due to the connection with black holes in two
dimensions and the Liouville theory \cite{WITTEN}. In this note we
will show a close connection between the first approach and the latter one
for the Euclidean coset. By imposing conformal invariance of the complete
$SU(1,1)$ theory coupled to an arbitrary unitary conformal field theory
(UCFT), we will show that the resulting physical state space will
automatically also satisfy the Euclidean coset condition, when we
restrict ourselves to the representations which have been proved to be
unitary for the $SU(1,1)$ string theory \cite{HWANG}. Therefore, in
these cases the physical properties of the $SU(1,1)$ string theories are
basically those of the coset.
This does not
mean that the two theories are equivalent. A string theory based on the coset
coupled with a UCFT will have a mass shell condition $L_0-1=0$, which
depends on the eigenvalue of $J^3_0$, the global generator in the compact
direction. This is important, since it is this dependence, which makes $L_0$
of the coset have only positive eigenvalues in the unitarity region. Therefore,
the mass-shell condition will only allow ground states.
This is in
contrast to the full $SU(1,1)$ theory, which has, in general, many possible
excitation levels. Outside the unitarity region, however, the coset model may
exhibit a richer structure.

Our results here also include an
alternative and more direct proof of unitarity of the $SU(1,1)$ string theory
than the one given in ref.\cite{HWANG}. It is more suitable for generalizations
to constructions using other non-compact groups.

Let $L_n^{(0)}$ be the $SU(1,1)$ Virasoro generators
\begin{equation}L_n^{(0)}\equiv {1\over x+2}\sum_m g_{ab}:J_m^aJ_{n-m}^b:~
.\end{equation} The total Virasoro generators are given by
$L_n=L_n^{(0)}+L_n^\prime$ where $L_n^\prime$ represent any UCFT. Here the
current modes $J^a_n$,  $a=1,2,3$, satisfy the affine $s\hat u(1,1)$ algebra
\begin{equation}[J^a_m,J^b_n]=if^{ab}_{~~c}J^c_{m+n}+{x\over
2}mg^{ab}\delta_{m+n}~, \end{equation}
with $g^{ab}=diag(-1,-1,1)$ (see \cite{HWANG} for further
conventions). We will also define the Virasoro generators of the
$SU(1,1)/U(1)$ coset, $L^{(3)}_n\equiv
L_n^{(0)}-{1\over x}\sum_m:J^3_{n-m}J^3_m:$. This coset
is defined by the requirement that $J_n^3, n>0$, annihilates
states. $L_n$ satisfy the
Virasoro algebra with conformal anomaly $c={3x\over x+2}+c^\prime$,
where $c^\prime$ is the conformal anomaly of $L_n^\prime$. We will here
always assume the critical value $c=26$ and $x<-2$, so that for the conformal
anomaly of the  $SU(1,1)$ theory, $c^{(0)}>3$. Physical states will satisfy
\begin{equation}L_n\ket{\psi}=(L_0-1)\ket{\psi}=0,~~~~n>0.\label{iii}
\end{equation}

Let us now
investigate these equations closer and show that in addition to the
Virasoro conditions, the states $\ket{\psi}$ will automatically satisfy
further conditions. We will first show the following property (I):
{\sl The following states
\begin{equation}\ket{\{\lambda\}\{\mu\}}=(J_{-1}^3)^{\lambda_1}\ldots
(J_{-N}^3)^{\lambda_N}L_{-1}^{\mu_1}\ldots
L_{-N}^{\mu_N}\ket{t;h,m}\label{iv}\end{equation}
are all linearly independent, when
$\tilde{D}^{N}_{vir}(h-{m^2\over x})\neq 0$.} Here
$L_n\ket{t;h,m}=J^3_n\ket{t;h,m}=0,~n>0$, and $h$ and $m$ are the eigenvalues
of
$L_0$ and $J^3_0$, respectively. $\tilde{D}^{N}_{vir}(\tilde{h})$ is the
determinant at level $N$ of the $\tilde{L}_n$-module ($\tilde{L}_n\equiv
L_n^{(3)}+L_n^\prime$) over a highest weight $\tilde{h}$.

To show this we first note that by using
$L_n=\tilde{L}_n+{1\over x}\sum_m:J^3_{n-m}J^3_m:$,
we can eliminate all factors of
$L_n$ in eq. \ref{iv} in favour of $\tilde{L}_n$. Then we consider the matrix
\begin{eqnarray}{\cal
M}^N_{\{\lambda^\prime,\mu^\prime\}\{\lambda,\mu\}}(h,m)\equiv &
\bra{t;h,m}(\tilde{L}_N)^{\mu_N^\prime}\ldots (\tilde{L}_1)^{\mu^\prime_1}
(J_N^3)^{\lambda_N^\prime}\ldots
(J_1^3)^{\lambda_1^\prime}\nonumber\\&(J^3_{-1}) ^{\lambda_1}\ldots
(J^3_{-N})^{\lambda_N}(\tilde{L}_{-1})^{\mu_1}\ldots
(\tilde{L}_{-N})^{\mu_N}\ket{t;h,m} .\end{eqnarray}
 If this matrix has a
non-vanishing determinant (for $\VEV{t\vert t}\neq 0$) then we have proved the
linear independence of the states in eq. \ref{iv}. This determinant is,
however,
easily calculated since $J^3_n$ and $\tilde{L}_n$ commute. It is, therefore,
given as a product of three factors. One from the tranverse space,
$\ket{t}$, the second one from the $U(1)$ space spanned by $J^3_n$ and
thirdly determinants of the $\tilde{L}_n$-modules up to level $N$ with a
highest weight of dimension $h-{m^2\over x}$. The first two factors are
non-zero and, therefore, the zeros of the $det\cal{M}$ come from the third
factor.This last factor is non-zero if the determinant of the Verma module at
level $N$ is non-zero and then the linear independence holds.

We can now use results established in the old proof of the
no-ghost theorem of the bosonic string \cite{GODDARD}. We
have the following property (II): {\sl All spurious states, \ie states of the
form $L_{-n}\ket{\chi}, n>0$, which are on-shell, are mapped to spurious states
under the action of $L_n, n>0$, if $c=26$. The space of non-spurious states is
invariant under the action of $L_n, n\geq 0$.} We will not prove the statement
here, but refer to the original proof in \cite{GODDARD}. With the basis eq.
\ref{iv} we write a general on-shell state $\ket{\psi}$ as
$\ket{\psi}=\ket{\phi}+\ket{s}$, where $\ket{s}$ is a spurious state.
Then by property II, we find that the equation $L_n\ket{\psi}=0, n>0$,
implies that $L_n\ket{\phi}=L_n\ket{s}=0, n>0$. The first of these
equations will imply, by using $L_n=\tilde{L}_n+{1\over
x}\sum_m:J^3_{n-m}J^3_m:$,  that $\sum_m:J^3_{n-m}J^3_m:\ket{\phi}=0$, $n>0$.
Consequently we must have $\ket{\phi}=\ket{t}$, so that the general
solution to eq. \ref{iii} will be $\ket{\psi}=\ket{t}+\ket{s}$. The spurious
states are null and the relevant part of the physical state space is given
by the transverse space $\ket{t}$. These states satisfy the additional
condition that they are annihilated by $J^3_n, n>0$.

We must now investigate under what circumstances the Kac-determinant of
$\tilde{L}_n$ will be non-zero. For the continous unitary representations the
situation is quite trivial, since the mass-shell condition will not allow any
excitations beyond the groundstates. Let us, therefore, consider the discrete
representations. The eigenvalues of $\tilde{L}_0$, $\tilde{h}$, are of the
form $\tilde{h}={j(j+1)\over x+2}-{m^2\over x}+h^\prime+N$, where $N$ is
the level of the state and $h^\prime$ the ground-state eigenvalue of
$L^\prime_0$ ($h^\prime\geq 0$ from unitarity). We can rewrite this as
\begin{equation}\tilde{h}={j(x-2j)\over
x(x+2)}+{1\over x}N(x-2j)+{1\over x}2j(j+N\pm m)-{1\over x}(j\pm
m)^2+h^\prime\end{equation} For the discrete representations $D_j^\pm$ we have
$j<0$ and $j\pm m\geq 0$, so that $\tilde{h}>0$ for $j>\frac{x}{2}$. From the
form of the Kac-determinant it is known that for the values of $\tilde{c}$
considered here, $2<\tilde{c}\leq 25$ there will be no zeros for positive
highest weights. This means that there will be no zeros for $j>{x\over 2}$.
Outside this region there will exist null-states, the first one for $j={x\over
2}$: $\tilde{L}_{-1}\ket{0;j=\mp m={x\over 2};h^\prime=0}$. The states  for
which
$j>{x\over 2}$ are precisely those which have been proved to be unitary for
the coset \cite{DIXON} and, consequently, we must also have unitarity for the
full $SU(1,1)$ string in this regime. We can summarize our results in the
following

\noindent{\bf Theorem}. {\sl The solutions of the Virasoro conditions,
with $j>{x\over 2}$} for the discrete representations,
will be unitary states $\ket{\phi}$, which are determined up to null-states and
will satisfy $$J^3_n\ket{\phi}=0, n>0.$$

One should compare this result with the proof of the no-ghost theorem given
previously \cite{HWANG}. In that proof one uses the light-like
combination $K_n=k\cdot J_n$ instead of $J^3_n$, where $k^a$ is light-like
$SU(1,1)$ vector. It is then established that physical states will be
annihilated by positive modes of $K_n$ provided the eigenvalues of $K_0$
are non-zero, {\it irrespective} of the value of $j$. This means that
the non-unitary representations will also be possible. In the case of
the time-like coset with $J^3_n$, unitarity appears to be manifest.
One should bear in mind, however, that the techniques used above
only provide sufficient conditions. It may still be that the physical states
satisfy the coset conditions even for representations which can have zeros in
the Kac determinant.

A simple application of the methods used
above, is the ordinary bosonic string in $d$-dimensional Minkowski space
coupled
to an arbitrary UCFT. We can copy the analysis using now the Virasoro
generators
$\tilde{L}_n\equiv L_n+{1\over 2}\sum_m\alpha_{n-m}^0\alpha_m^0$ and in place
of
$J_n^3$ we use $\alpha^0_n$. Then, if the determinant of the corresponding
Verma module is non-zero, the physical states will satisfy
$\alpha^0_n\ket{\phi}=0, n>0$. The eigenvalues of $\tilde{L}_0$ are strictly
positive except when all of the space-like components of the momentum
$p^i$, $i=1\ldots d-1$, are zero. But, by Lorentz invariance, we can, at any
mass
level, choose a frame for which $p^i\neq 0$, unless $p^\mu\equiv 0$ or $d<2$.
  Consequently, for strings in dimensions $2\leq d \leq 26$ and non-zero
momentum we will have no  Virasoro null states and the time-like
condition is implied. The unitarity of the physical subspace is then manifest.
This means that the time-like gauge is possible for the quantum theory for
$2\leq
d \leq 26$.

As we have mentioned, there exists another coset space, which is possible as
a solution to the Virasoro conditions. This space was specified by the
light-like combination $K_n=J^1_n+J^3_n$. The coset arises if we choose to work
with ground-states which are diagonalized with respect $K_0$. In the covariant
quantization it is, therefore, essentiallly the choice of the basis of the
ground-states which determines which gauge-slice we are to find. An interesting
question is what happens if we take the ground states in a
basis which is diagonal with respect to one of the non-compact generators \eg
$J^2_0$. The natural coset would then be defined by modding out $J^2_n$ and
corresponds to the Minkowskian version of the black hole in
ref.\cite{WITTEN}. In this case we have $\tilde{L}_n\equiv
L_n^{(2)}+L_n^\prime$,
$L^{(2)}_n\equiv L_n^{(0)}+{1\over x}\sum_m:J^2_{n-m}J^2_m:$. The important
difference here, as compared with the compact case, is that the eigenvalues of
$\tilde{L}_0$ are now of the form $\tilde{h}= {j(j+1)\over x+2}+{\lambda^2\over
x}+h^\prime+N$ ($\lambda$ is the eigenvalue of $J^2_0$, which is real for the
unitary representations). These values are in general both positive and
negative and consequently our techniques can not be used to show that this
coset
is contained in the Virasoro conditions. This is an indication that the coset
is not a good gauge slice. It is perhaps not surprising. In the limit of a
flat manifold, the coset would correspond to a gauge in which we
fix one of the space-like components of $\alpha_n^\mu$. This is classically not
an acceptable gauge. One may also check explicitely, that the solutions to
the Virasoro conditions at level two do not satisfy the space-like
coset condition, so that the coset is not a permissable gauge at the quantum
level, either. If this is still true for non-zero curvature, then the question
is
in which coset the physical states lie. Such a coset must exist, since by gauge
invariance the dimensionality of the physical state space is such that it
admits
subsidiary gauge conditions.

I would like to thank Lars Brink and Robert Marnelius for discussions.

\end{document}